\documentclass[english,12pt,preprint]{aastex}
\usepackage[T1]{fontenc}
\usepackage[latin1]{inputenc}
\makeatletter
\usepackage{babel}
\makeatother
\begin{document}

\title{Evaluating the Magnetic Field Strength in Molecular Clouds}

\author{Martin Houde}

\email{houde@astro.uwo.ca}

\affil{Department of Physics and Astronomy, University of Western Ontario,
London, Ontario, N6A 3K7, Canada}

\affil{Caltech Submillimeter Observatory, 111 Nowelo Street, Hilo, HI 96720}


\begin{abstract}
We discuss an extension to the Chandrasekhar-Fermi method for the
evaluation of the mean magnetic field strength in molecular clouds
to cases where the spatial orientation of the field is known. We apply
the results to M17, using previously published data.
\end{abstract}

\section{Introduction}

There exist few techniques that allow for the measurement of quantities
that characterize the magnetic field in molecular clouds. At millimeter
and submillimeter wavelengths, the orientation of the magnetic field
is most commonly traced using polarimetry measurements from dust continuum
emission \citep{Hildebrand 1988}. The strength of the magnetic field
(in general, its line-of-sight component) can only be directly measured
via the Zeeman effect (e.g., \citet{Crutcher 1999, Brogan 2001}),
usually at longer wavelengths. In order to gather as much information
as possible about the magnetic field, the so-called Chandrasekhar
Fermi (CF) method \citep{Chandra 1953} is often used to infer the
strength of the plane-of-the-sky component of the field. Because this
is achieved with the same polarimetry data that give the orientation
of the sky-projected magnetic field, the CF method can act as a bridge
between the polarimetry and Zeeman observations to provide an estimate
for the magnitude of the mean field strength in a given cloud. 

In this paper, we discuss how a simple extension of the CF method
can be used alone, i.e., without the need of Zeeman data, to infer
the magnitude of the magnetic field; not only the strength of its
plane-of-the-sky component. Furthermore, it will also be shown that,
contrary to the original CF method which only really works well when
the magnetic field is located close enough to the plane of the sky,
our generalization is valid regardless of the field's orientation
in space. However, this can only be accomplished if and when the spatial
orientation of the magnetic field is known. That is to say, not only
the orientation of its projection on the plane of the sky is needed
(from polarimetry), but also its inclination to the line of sight.
This last piece of information can be obtained through the technique
of \citet{Houde 2002} which relies on the availability of spectroscopic
measurements from suitable neutral and ionic molecular species, as
well as polarimetry.

Finally, we apply our extension to the CF method to already published
data for the M17 molecular cloud \citep{Houde 2002}, and infer a
value for the magnitude of the mean magnetic field for this object.

\section{The CF equation}

It was originally asserted by \citet{Chandra 1953} that the amount
of dispersion of the polarization angles measured from starlight (or
dust continuum radiation) can reveal information about the magnitude
of the magnetic field. With the assumption that the magnetic field
is frozen to the ambient fluid, any (turbulent) motion within the
gas in a direction perpendicular to the orientation of the magnetic
field will be transmitted to, and distort, the field lines. \citet{Chandra 1953}
further assumed that such disturbances would propagate as waves along
the magnetic field lines at the Alfv\'en speed, which they used as
the starting point for their analysis. It follows that since dust
grains are thought to be tied to the magnetic field lines \citep{Mouschovias 1999},
the amount of distortion in the field lines can be inferred from polarimetry.
Similarly, the turbulent motion of the gas can be measured through
the spectral line profiles of molecular species, for example. These
two observed quantities are needed to evaluate the strength of the
magnetic field through the CF method.

Following, therefore, the original derivation of \citet{Chandra 1953},
we can write an equation for the mean value of the magnetic field
$B$ as

\begin{equation}
B=\sqrt{4\pi\rho}\frac{\sigma\left(v_{\perp}\right)}{\sigma\left(\phi\right)},\label{eq:CF1}\end{equation}

\noindent where $\rho$ and $\sigma\left(v_{\perp}\right)$ are, respectively,
the mass density and the two-dimensional velocity dispersion (perpendicular
to the field lines) of the matter coupled to the magnetic field. $\sigma\left(\phi\right)$
is the dispersion in angular deviations of the field lines. Equation
(\ref{eq:CF1}) is valid in the small angle limit. 

\begin{figure}
\epsscale{0.8}

\plotone{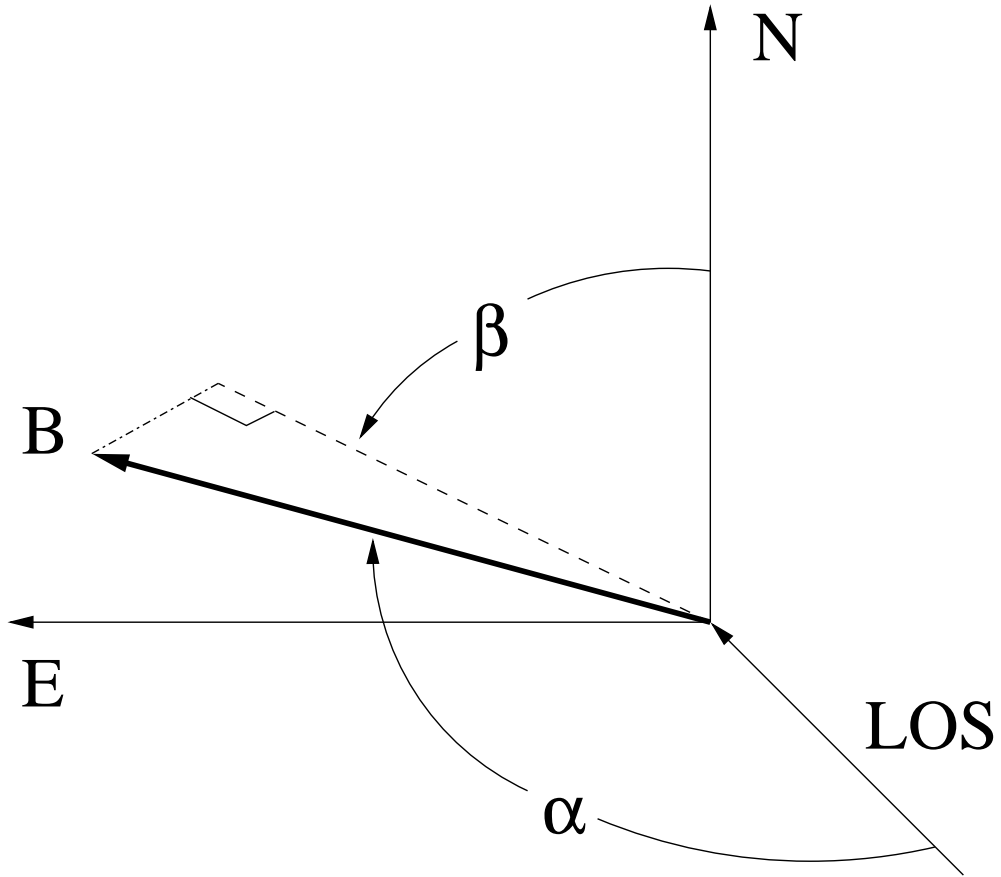}

\caption{\label{cap:coord}The spatial orientation of the magnetic field is
defined with the two angles $\alpha$ and $\beta$. The N, E, and
LOS axes stand for north, east, and line of sight, respectively. From
\citet{Houde 2002}.}
\end{figure}

\noindent In their estimation of the magnetic field strength in the
spiral arms \citet{Chandra 1953}, identified $\sigma\left(\phi\right)$
with the dispersion in the orientation of the polarization vectors
measured for distant background stars. Using the coordinate system
of Figure \ref{cap:coord} to define the spatial orientation of the
magnetic field, with $\alpha$ the inclination angle of the field
to the line of sight, and $\beta$ the angle made by its projection
on the plane of the sky, we find, for the case originally considered
by \citet{Chandra 1953}, that

\begin{equation}
\sigma\left(\phi\right)=\sigma\left(\beta\right).\label{eq:sig(b)}\end{equation}

However, observations of this type probe only one direction in the
lateral displacement of the magnetic field line. We must, therefore,
make the following substitution for the velocity dispersion

\begin{equation}
\sigma\left(v_{\perp}\right)\rightarrow\frac{1}{\sqrt{2}}\sigma\left(v_{\perp}\right)=\frac{1}{\sqrt{3}}\sigma\left(v\right),\label{eq:sigmas}\end{equation}

\noindent where $\sigma\left(v\right)$ is the total three-dimensional
velocity dispersion of the gas (for cases of isotropic turbulence).
Inserting equations (\ref{eq:sig(b)}) and (\ref{eq:sigmas}) in equation
(\ref{eq:CF1}) we obtain the original equation derived by \citet{Chandra 1953}

\begin{equation}
B_{{\rm {pos}}}=\sqrt{\frac{4}{3}\pi\rho}\frac{\sigma\left(v\right)}{\sigma\left(\beta\right)},\label{eq:CF_org}\end{equation}

\noindent where $B_{{\rm {pos}}}$ is the plane-of-the-sky component
of the magnetic field (more on this below).

Equation (\ref{eq:CF_org}) is often used to measure the mean strength
of the plane-of-the-sky component of the magnetic field in molecular
clouds (e.g., \citet{Lai 2003b}). It has also been tested with magnetohydrodynamic
(MHD) simulations to verify its domain of applicability \citep{Ostriker 2001, Padoan 2001, Heitsch 2001, Kudoh 2003}.
Although the CF method has been found to work well for strong enough
magnetic field, it also suffers from some shortcomings. Among these,
is the fact that the equation (\ref{eq:CF_org}) only really applies
well when the magnetic field is located close enough to the plane
of the sky. In fact, the method will fail when the field is aligned
parallel to the line of sight ($\alpha=0$ in Figure \ref{cap:coord}).

\subsection{An extension to the CF method}

It would be desirable to extend the CF method to cases where the magnetic
field is arbitrarily oriented in space. This, however, requires that
observations can be made to measure not only $\beta$ (the angle made
by the projection of the magnetic field on the plane of the sky),
but also $\alpha$ (the inclination angle of the field to the line
of sight). Some methods have already been proposed to do such measurements.
\citet{Myers 1991} (see also \citet{Bourke 2004}) modeled the magnetic
field in molecular clouds with uniform and nonuniform components,
and through a statistical analysis were able to evaluate the spatial
orientation (i.e., they inferred $\alpha$ and $\beta$) for the mean
three-dimensional uniform field. More recently, \citet{Houde 2002}
have proposed a technique that combines polarimetry and ion-to-neutral
line width ratio measurements \citep{Houde 2000a, Houde 2000b} to
map the spatial orientation of the magnetic field across molecular
clouds. This method has been used so far for three different objects:
M17 \citep{Houde 2002}, DR 21(OH) \citep{Lai 2003a}, and Orion A
\citep{Houde 2004}.

Once $\alpha$ and $\beta$ are mapped across a given molecular cloud,
the angular dispersions $\sigma\left(\alpha\right)$ and $\sigma\left(\beta\right)$
can be calculated from the measured data. It is easy to show that,
in the small angle limit, the total angular dispersion of the magnetic
field lines $\sigma\left(\phi\right)$ is given by

\begin{equation}
\sigma^{2}\left(\phi\right)=\sigma^{2}\left(\alpha\right)+\sin^{2}\left(\alpha\right)\sigma^{2}\left(\beta\right).\label{eq:sigma(phi)}\end{equation}

Equation (\ref{eq:sigma(phi)}) takes into account not only the inclination
of the magnetic field, but also angular deviations along two independent
directions perpendicular to the field orientation. Because of this
last point, the velocity dispersion will be $\sqrt{2}$ times larger
than what is used in the original CF method equation (\ref{eq:CF_org}).
That is to say, we will now either use the two-dimensional velocity
dispersion $\sigma\left(v_{\perp}\right)$, define after equation
(\ref{eq:CF1}), or its equivalent expressed as a function of $\sigma\left(v\right)$
if the turbulence is isotropic 

\begin{equation}
\sigma\left(v_{\perp}\right)=\sqrt{\frac{2}{3}}\sigma\left(v\right).\label{eq:sigma(per)}\end{equation}

Using equations (\ref{eq:sigma(phi)}) and (\ref{eq:sigma(per)}),
we can now write a generalized CF equation from (\ref{eq:CF1})

\begin{equation}
B=C\left[\frac{4\pi\rho\sigma^{2}\left(v_{\perp}\right)}{\sigma^{2}\left(\alpha\right)+\sin^{2}\left(\alpha\right)\sigma^{2}\left(\beta\right)}\right]^{\frac{1}{2}},\label{eq:CF_mod}\end{equation}

\noindent or if the turbulence is isotropic

\begin{equation}
B=C\left[\frac{8\pi\rho\sigma^{2}\left(v\right)}{3\left[\sigma^{2}\left(\alpha\right)+\sin^{2}\left(\alpha\right)\sigma^{2}\left(\beta\right)\right]}\right]^{\frac{1}{2}}.\label{eq:CF_iso}\end{equation}

In both equations (\ref{eq:CF_mod}) and (\ref{eq:CF_iso}) we have
added a correction factor $C$ (first introduced by \citet{Ostriker 2001})
to take into account some shortcomings of the CF method to be discussed
later. It is now easy to see how equation (\ref{eq:CF_iso}) can be
reduced to one for the plane-of-the-sky component of the magnetic
field $B_{{\rm {pos}}}$ (i.e., equation (\ref{eq:CF_org})) when
only polarization measurements are available. In this case, for a
sufficiently large set of data we expect (as long as $\alpha\neq0$)

\[
\sigma^{2}\left(\alpha\right)=\sin^{2}\left(\alpha\right)\sigma^{2}\left(\beta\right),\]

\noindent and

\[
\sigma^{2}\left(\phi\right)=2\sin^{2}\left(\alpha\right)\sigma^{2}\left(\beta\right).\]

\noindent We can write

\begin{eqnarray*}
B_{{\rm {pos}}} & = & B\sin\left(\alpha\right)\\
 & = & \sqrt{\frac{4}{3}\pi\rho}\frac{\sigma\left(v\right)}{\sigma\left(\beta\right)},\end{eqnarray*}

\noindent which is the same as equation (\ref{eq:CF_org}).

We can, therefore, emphasize two important advantages of the modified
CF equation (\ref{eq:CF_mod}) (or (\ref{eq:CF_iso})) over the original:

\begin{itemize}
\item the new equation is valid no matter what the orientation of the magnetic
field is. Most notably, the method does not fail when the field is
directed along the line of sight. 
\item Finally, the value for the magnetic field calculated with equation
(\ref{eq:CF_mod}) is not that of its plane-of-the-sky component,
but that of \emph{full magnitude of the mean magnetic field vector}.
\end{itemize}

\subsection{Shortcomings of the method\label{sub:coupling}}

As mentioned earlier, MHD simulations have already been used in the
past \citep{Ostriker 2001, Padoan 2001, Heitsch 2001, Kudoh 2003}
to test the validity of the original CF method (equation (\ref{eq:CF_org})).
The main conclusion of these studies was that the introduction of
a correction factor ($C$ in equations (\ref{eq:CF_mod}) and (\ref{eq:CF_iso}))
is needed when evaluating $B_{{\rm {pos}}}$. A correction of $C\sim0.5$
was deemed appropriate in most cases when the field is not too weak.
A few reasons are usually identified for this. For example:

\begin{enumerate}
\item Smoothing of the field: because of the finite resolution with which
observations are done, there will be an averaging of the angular structure
of the field. This will bring a decrease of the angular dispersion
$\sigma\left(\phi\right)$, and an overestimation of the field strength
\citep{Ostriker 2001}.
\item Similarly, line-of-sight averaging (independent of the angular resolution
of the observations) of the magnetic field will decrease $\sigma\left(\phi\right)$
\citep{Myers 1991}.
\item Inhomogeneity and complex density structures (e.g., clumpiness) also
tend to reduce the value of $C$ \citep{Zweibel 1990, Ostriker 2001}.
\end{enumerate}
We also add to the previous points one more aspect that should be
kept in mind when applying the CF method.

In the case of highly turbulent and massive molecular clouds (like
in the example considered in the next section), it has been observed
that there can exist significant velocity drifts between coexistent
neutral and ionic molecular species. This can be ascertained through
the comparison of the observed line profiles for the two types of
species, the ions consistently exhibiting narrower spectral line widths
\citep{Houde 2000a, Houde 2000b, Houde 2002, Lai 2003a, Houde 2004}.
This implies that the coupling between ions and neutrals is not perfect
\citep{Houde 2002}. Within the context of the CF method, this bring
about uncertainties in two of the quantities used when evaluating
the magnetic field strength. Indeed, because of this imperfect coupling
between ions and neutrals, the mass density $\rho$ used in the CF
equation cannot be that of (larger) neutral density. It must be somewhat
smaller. Furthermore, because of the aforementioned velocity drift,
the velocity dispersion perpendicular to the field lines $\sigma\left(v_{\perp}\right)$
(or $\sigma\left(v\right)$) cannot be that measured for a neutral
molecular species. It must also be smaller. The combination of these
factors will also tend to reduce the value of $C$ (in equations (\ref{eq:CF_mod})
or (\ref{eq:CF_iso})), at least when the CF method is applied to
highly turbulent and massive molecular clouds.

We leave the quantification of these effects as open questions that
could, perhaps, be investigated through simulations.

\section{Application of the extended CF method to M17}

Using their aforementioned technique, \citet{Houde 2002}  measured
the spatial orientation of the magnetic field at 57 different positions
across the M17 molecular clouds. This was accomplished using extensive
350 $\mu$m dust continuum polarimetry and spectroscopy (HCO$^{+}$/HCN)
maps obtained at the Caltech Submillimeter Observatory. We now use
their results to calculate mean magnetic field strength for M17, using
equation (\ref{eq:CF_mod})%
\footnote{The values for $\alpha$ and $\beta$ used here are slightly different
from those presented in \citet{Houde 2002}. We use a maximum polarization
level of 10\%, instead of 7\% as was used in their original analysis.
See \citet{Houde 2004} for more details.%
}.

From the analysis of \citet{Houde 2002} we find the following averages
for M17:

\begin{eqnarray*}
\alpha & \simeq & 47.1^{\circ},\\
\sigma\left(\alpha\right) & \simeq & 10.8^{\circ},\\
\beta & \simeq & 76.1^{\circ},\\
\sigma\left(\beta\right) & \simeq & 16.7^{\circ}\\
\sigma\left(\phi\right) & \simeq & 16.3^{\circ},\\
\sigma\left(v_{\perp}\right) & \simeq & 2.0\,{\rm {km/s}.}\end{eqnarray*}

The transverse velocity dispersion was evaluated from the HCN spectra,
taking into account the (anisotropic) turbulent flow model used by
\citet{Houde 2002} (see their Figure 2, and equation (11)), and the
fact that the inclination angle is known%
\footnote{Within the context of the anisotropic turbulent model of \citet{Houde 2002},
a value for $\sigma\left(v_{\perp}\right)$ at each position can be
obtained from the corresponding observed spectral line width $\sigma_{{\rm {obs}}}\left(v\right)$.
It can be shown that $\sigma^{2}\left(v_{\perp}\right)=\sigma_{{\rm {obs}}}^{2}\left(v\right)f/\left[e\cos^{2}\left(\alpha\right)+f/2\,\sin^{2}\left(\alpha\right)\right]$,
where $e$ and $f$ are given in their equation (11) with $\Delta\theta=44.4^{\circ}$.%
}. Upon using equation (\ref{eq:CF_mod}) with $C=0.5$, an approximate
value of $10^{6}$ cm$^{-3}$ for the mean density, and a mean molecular
mass of 2.3, we find

\[
B\approx2.5\,{\rm {mG}.}\]

This value for the magnitude of the magnetic field could be further
reduced if the correction factor $C$ were found to be smaller than
the stated value (because of the effects discussed in section \ref{sub:coupling}),
or again if the average density across the maps were less than what
was assumed here. However, this field strength may not be too excessive
in light of the fact that \citet{Brogan 2001} obtained a peak value
of $-750\,\mu$G for the line-of-sight component of the magnetic field
in M17, using HI Zeeman measurements. For, once the inclination angle
quoted above is taken into account, we calculate from their data a
field magnitude in excess of 1 mG. Our molecular species (i.e., HCN
and HCO$^{+}$, in the $J=4\rightarrow3$ transition) probe denser
media which could harbor stronger fields.

It is also interesting to note that

\[
\sigma\left(\alpha\right)\sim\sin\left(\alpha\right)\sigma\left(\beta\right)=12.2^{\circ},\]

\noindent as would be expected for a large enough data set.

Finally, we would like to state that the extension to the CF method
presented in this paper should be readily testable through MHD simulations,
as was done in the past for the original CF technique \citep{Ostriker 2001, Padoan 2001, Heitsch 2001, Kudoh 2003}.

\acknowledgements{The author thanks T.G. Phillips, R. Peng, and S. Basu for helpful
discussions. The Caltech Submillimeter Observatory is funded by the
NSF through contract AST 99-80846.}

\end{document}